\documentclass[conference]{IEEEtran}
\IEEEoverridecommandlockouts
\usepackage{cite}
\usepackage{amsmath,amssymb,amsfonts}
\usepackage{graphicx}
\usepackage{textcomp}
\usepackage{xcolor}
\usepackage{multirow}

\usepackage[ruled,vlined,linesnumbered]{algorithm2e}

\def\BibTeX{{\rm B\kern-.05em{\sc i\kern-.025em b}\kern-.08em
    T\kern-.1667em\lower.7ex\hbox{E}\kern-.125emX}}

\begin{document}

\title{Targeted Augmented Data for Audio Deepfake Detection\\
\thanks{This work was supported by the Luxembourg National Research Fund (FNR) under the project BRIDGES2021/IS/16353350/FaKeDeTeR and POST Luxembourg.}
}

\author{\IEEEauthorblockN{Marcella Astrid\IEEEauthorrefmark{1} \qquad
Enjie Ghorbel\IEEEauthorrefmark{1}\IEEEauthorrefmark{2} \qquad Djamila Aouada\IEEEauthorrefmark{1}}
\IEEEauthorblockA{\IEEEauthorrefmark{1}Interdisciplinary Centre for Security, Reliability and Trust (SnT), University of Luxembourg, Luxembourg \\
\IEEEauthorrefmark{2}Cristal Laboratory, National School of Computer Sciences (ENSI), Manouba University, Tunisia \\
Email: marcella.astrid@uni.lu,
enjie.ghorbel@isamm.uma.tn,
djamila.aouada@uni.lu}}

\maketitle

\begin{abstract}
The availability of highly convincing audio deepfake generators highlights the need for designing robust audio deepfake detectors. Existing works often rely solely on real and fake data available in the training set, which may lead to overfitting, thereby reducing the robustness to unseen manipulations. To enhance the generalization capabilities of audio deepfake detectors, we propose a novel augmentation method for generating audio pseudo-fakes targeting the decision boundary of the model. Inspired by adversarial attacks, we perturb original real data to synthesize pseudo-fakes with ambiguous prediction probabilities. 
Comprehensive experiments on two well-known architectures demonstrate that the proposed augmentation contributes to improving the generalization capabilities of these architectures.

% The surge in highly convincing fake audio content due to advancements in generative artificial intelligence necessitates robust audio deepfake detectors to combat misinformation and fraudulent activities. While prior research has predominantly concentrated on architecture design, we advocate for a shift towards data augmentation to enrich the diversity of fake data. Inspired by recent works in anomaly detection, we propose targeting the creation of pseudo-fake data at the decision boundary between real and fake. By employing insights from adversarial attacks, we perturb input data during training to generate augmented fake data strategically positioned near this boundary. This approach not only effectively encapsulates real data but also encourages diversity through evolving decision boundaries. Our method, implemented on state-of-the-art audio deepfake architectures, showcases improved performance compared to models trained without augmented data. Through comparisons with untargeted augmentation and targeted augmentation solely at fake data, we validate the efficacy of our approach in enhancing model performance. 
% Notably, our method exhibits robustness across different hyperparameter settings and competes effectively against existing techniques, offering a promising direction for enhancing audio deepfake detection capabilities.
\end{abstract}


\section{Introduction}

With the rapid advancement of generative deep learning, creating extremely convincing fake audio content, known as audio deepfakes, has become feasible. While such progress offers numerous advantages, it also opens avenues for malicious exploitation, such as spreading false information~\cite{biden-fake} or facilitating scams~\cite{scam}. Consequently, introducing methods for detecting audio deepfakes is crucial.

\begin{figure}[]
  \centering
  \includegraphics[width=\linewidth]{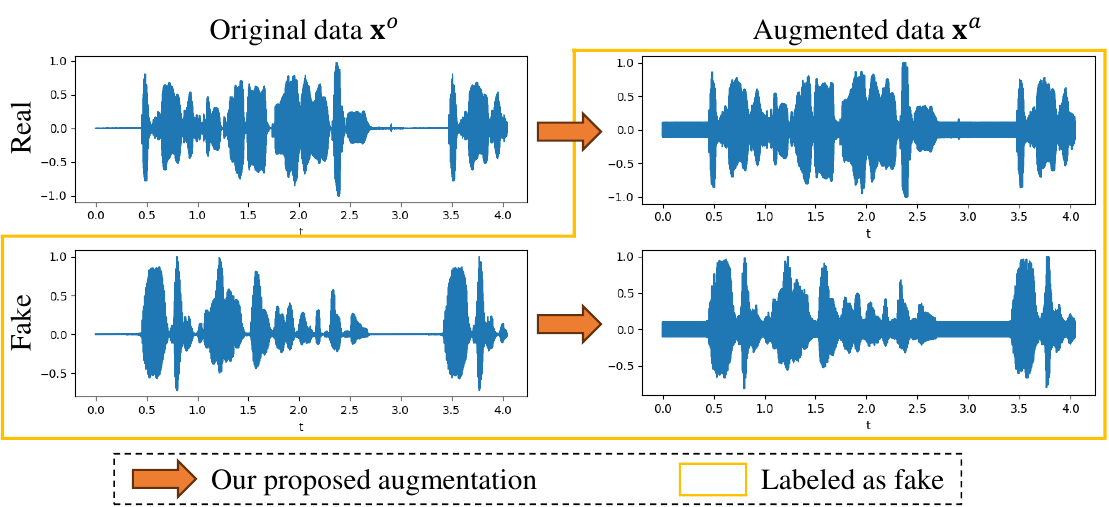}
  \vspace{-7mm}
  \caption{Overview of the proposed augmentation method for increasing the diversity of fake data.
  % Original dataset data is augmented to enhance the diversity of fake data. 
  Both the original fake data and the augmented data are labeled as fake. The combined dataset is utilized during training. The x-axis denotes time, while the y-axis represents the audio signal magnitude.}
  \vspace{-5mm}
  \label{fig:introduction}
\end{figure}

Previous works on audio deepfake detectors have primarily focused on the architecture design, either through manual design~\cite{jung2022aasist,tak2021endgraph,tak2021end} or neural architecture search~\cite{ge2021partially,ge2021raw}. The training process typically involves the use of both real and fake data provided in the dataset. However, as for any deep learning-based method, such an approach tends to overfit the manipulation methods seen during the training phase, hence leading to poor performance when encountering unseen manipulations. %variety of fake data is theoretically unlimited; for instance, future methods of generating deepfakes may diverge from those currently known. 
Therefore, in this study, we propose a shift in focus from architecture to data augmentation. To enhance the robustness to unseen deepfakes, we aim to increase the diversity of fake data by augmenting the existing training dataset, as illustrated in Figure~\ref{fig:introduction}.

 \begin{figure*}[]
  \centering
  \includegraphics[width=\linewidth]{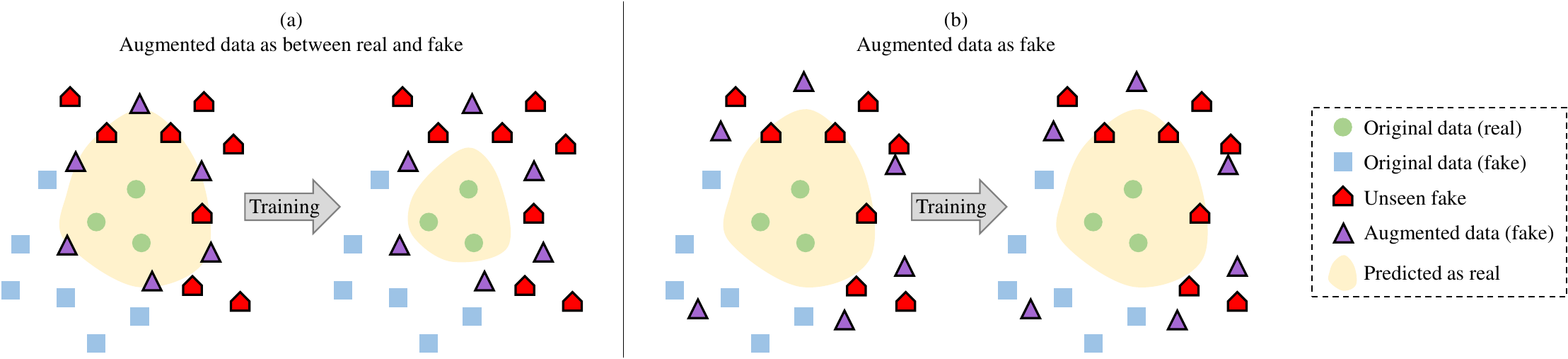}
  \vspace{-7mm}
  \caption{Illustration of two different augmentation strategies: (a) augmentation of fake data in the neighborhood of the decision boundary (between real and fake data); and (b) augmentation in the neighborhood of fake data without considering the decision boundary. 
  %As compared to (b), the strategy employed in (a) allows a better estimation of the decision boundary.
  As compared to (b), the strategy employed in (a) can better enhance the generalization to unseen fake.}
  \label{fig:method_intuition}
  \vspace{-5mm}
\end{figure*}

Specifically, we take inspiration from previous works on anomaly detection with a closely related objective, namely, generalizing to unseen anomalies. In fact, such an objective can be approached by generating pseudo-anomalies (i.e., augmented anomalous data, distinct from actual anomalous data) for covering more variability in anomalous data~\cite{zaheer2020old,astrid2021learning,ngo2019fence,dionelis2020tail,dionelis2020boundary}. 
% We draw inspiration from the concept of generating pseudo-anomalies (i.e., augmented fake data, distinct from actual fake data) in the anomaly detection domain to enhance our fake data. 
% % Anomaly detection shares similarities with deepfakes, as anomalies are theoretically unlimited. 
% Training with pseudo-anomalies aids the model in recognizing unseen anomalies during testing~\cite{zaheer2020old,astrid2021learning}, which purpose is similar to our purpose of detecting unseen manipulations. 
However, different pseudo-anomalies do not contribute equally to the enhancement in performance. In particular, several research works~\cite{ngo2019fence,dionelis2020tail,dionelis2020boundary} suggest that pseudo-anomalies located near the decision boundary separating normal data from fake ones can improve the model performance. 

In this paper, we argue that following a similar boundary-driven data augmentation strategy in the context of audio deepfake detection can be a way to enhance the classification performance.  To achieve this, we adopt a method from adversarial attacks~\cite{goodfellow2015explaining} and utilize the model under training to perturb the input, resulting in augmented fake data. 
In adversarial attacks, the objective is to perturb the input in a way that induces the model to confidently make an incorrect prediction. For instance, in the context of deepfake detection, we might perturb a fake image so that the model confidently and inaccurately predicts it as real. In contrast, we perturb the input such that the model predicts it as a mixture of real and fake by minimizing the loss towards a prediction that is half real and half fake.
%However, unlike conventional adversarial attacks that perturb inputs to target a wrong class prediction based on ground truth or maximize the loss, our approach perturbs towards a prediction that is half fake and half real. 
This guides the perturbed input toward the decision boundary of the model. 
This strategy offers two key advantages: Firstly, it positions the augmented data near the boundary of real data, thereby encapsulating the genuine data more effectively (as depicted in Figure~\ref{fig:method_intuition}). 
Secondly, since we continuously generate the augmented data throughout the training, the augmented data also evolves throughout epochs, introducing further diversity due to shifts in the decision boundary of the model. 
Our method focuses solely on the training data and remains architecture-agnostic. We apply this augmentation technique to train two state-of-the-art audio deepfake architectures, namely, AASIST~\cite{jung2022aasist} and RawNet2~\cite{tak2021end}. The reported experiments show that the proposed adversarial augmentation technique improves the performance of both architectures, thereby highlighting its relevance. 
% Moreover, the obtained results demonstrate that the proposed boundary-driven augmentation outperforms standard data augmentation technique based on Gaussian noise.

% It can also be noted that generating fake data from original dataset samples has also been explored in visual deepfake detection domain. Some methods~\cite{li2020face,shiohara2022detecting,mejri2023untag} employ handcrafted augmentation techniques, which are not adaptable during training, potentially limiting model optimization. Therefore, SLADD~\cite{chen2022self} suggests using a generator to determine the type of the generated pseudo-fake by maximizing the loss of the deepfake detector. In comparison, our method targets the boundary between real and fake to create fake data. Additionally, while SLADD trains its generator using reinforcement learning to approximate the gradient from the deepfake detector, potentially leading to training convergence issues, we directly utilize the gradient from the deepfake detector. Our comparisons with untargeted augmented data and data targeted solely at fake predictions can provide direct insights into handcrafted augmentation and vanilla adversarial training-based augmentation concepts.

In summary, our contributions can be summarized as follows: 1) We are, to the best of our knowledge, the first in audio deepfake detection to focus on increasing the variety of fake data through augmentation; 2) We propose an adversarial-attack-like method to target the augmented data toward the decision boundary between real and fake; 3) Extensive experiments show the effectiveness of the targeted augmentation to improve the performance of audio deepfake detectors.
% Our method which is architecture-agnostic is used to train two state-of-the-art architecture: RawNet2~\cite{tak2021end} and AASIST~\cite{jung2022aasist}; 4) To further verify our method, we compare it to untargeted augmented data with Gaussian noise and to targeted augmented data directed at fake rather than half-real-half-fake combinations; 5) Our method demonstrates robustness to changes in hyperparameters and is competitive to SoA techniques.

% We decide to focus on data and increasing variety in the fake data can be good. Therefore, we propose to augment data to create more fake data from the data in the dataset as depicted in Figure \red{[samples figure]}. Since augmented data has different pattern than the real original data, we set the augmented data label as fake. Then we train the model with both original data and augmented data.

% Inspiration from anomaly detection, create pseudo-fake near the boundary between real and fake. The way we augment our data is by targeted data to between real and fake throughout the training process. This gives 2 benefits: 1) the data is near the real data boundary which can encapsulate the real data more tight (the intuition is illustrated in Figure \red{[figure illustration]}); 2) the augmented data change over the training process which gives more variety in the overall training process due to change of decision boundary. it's generic: we experiment on 2 sota architectures: aasist and rawnet2.

\noindent\textbf{Paper organization:} We discuss related work on deepfake detection in Section~\ref{sec:related_work}. We detail our method in Section~\ref{sec:methodology}. Section~\ref{sec:experiments} covers the experimental setup and results. Finally, our conclusions are presented in Section~\ref{sec:conclusion}.

\section{Related Work}
\label{sec:related_work}
Audio deepfake detectors primarily focus on architectural designs, such as using smaller networks~\cite{hua2021towards,ma2021improved}, multiple feature scales~\cite{li2021replay,li2021channel}, fixed bandwidth filters~\cite{tak2021end}, attention mechanisms~\cite{tak2021graph,tak2021end,jung2022aasist,zhang122021effect}, and graph neural networks~\cite{tak2021graph,tak2021end,jung2022aasist}. Another approach is to automatically search for optimal architectures through neural architecture search techniques~\cite{ge2021partially,ge2021raw}. However, instead of focusing on the architectural design, we investigate the enrichment of training data by generating fake data from the original dataset.

While data synthesis strategies, to the best of our knowledge, have not been explored in the context of audio deepfake detection, generating pseudo-fake data from original dataset samples has been largely investigated in visual deepfake detection domain. Some methods~\cite{li2020face,shiohara2022detecting,mejri2023untag,nguyen2024laa} employ handcrafted augmentation techniques, which are fully decoupled from the training phase, potentially leading to sub-optimal augmentations. Therefore, SLADD~\cite{chen2022self} suggests using a generator to determine the type of the generated pseudo-fake by maximizing the loss of the deepfake detector. In comparison, our method targets the boundary between real and fake to create fake data. Additionally, while SLADD trains its generator using reinforcement learning to approximate the gradient from the deepfake detector, potentially resulting in training convergence issues, we directly utilize the gradient from the deepfake detector.
% Our comparisons with untargeted augmented data and data targeted solely at fake predictions can provide direct insights into handcrafted augmentation and vanilla adversarial training-based augmentation concepts.

% \subsection{Audio Deepfake Detectors}
% Various audio deepfake detector......
% However, all of the methods are training with real and fake available in the training set. We propose to generate more fake data with target to between real and fake boundary. Moreover, our method in principal is generic to these method.  

% \subsection{Augmentations}
% Normal augmentation. Ours is different as we consider the augmented data as fake because the pattern can be considered different from real data.

% Bad data augmentation in various domains. Bad GAN, Fence GAN. Ours make use the model itself without the necessities or dependencies of extra model.

\section{Methodology}
\label{sec:methodology}

% An audio detection model $D(\cdot)$ takes a batch of input $\mathbf{X} = \{\mathbf{x}_1, \mathbf{x}_2, ..., \mathbf{x}_B\}$ and outputs predictions $\hat{\mathbf{Y}} = \{\hat{\mathbf{y}}_1, \hat{\mathbf{y}}_2, ..., \hat{\mathbf{y}}_B\}$, where $B$ is the batch size, as follows
% \begin{equation}
%     \hat{\mathbf{Y}} = D(\mathbf{X})\text{.}
% \end{equation}
% The model is trained to minimize the classification loss $\mathcal{L}(\hat{\mathbf{Y}}, \mathbf{Y})$ between the prediction $\hat{\mathbf{Y}}$ and the label $\mathbf{Y} = \{\mathbf{y}_1, \mathbf{y}_2, ..., \mathbf{y}_B\}$. 

An audio detection model, denoted as $D(\cdot)$, processes an input $\mathbf{x}$
% $\mathbf{X} = \{\mathbf{x}_1, \mathbf{x}_2, ..., \mathbf{x}_B\}$ 
and generates the prediction denoted as $\hat{\mathbf{y}}$.
% $\hat{\mathbf{Y}} = \{\hat{\mathbf{y}}_1, \hat{\mathbf{y}}_2, ..., \hat{\mathbf{y}}_B\}$, where $B$ 
% represents the batch size. 
Mathematically, this can be expressed as follows,
\begin{equation}
    \hat{\mathbf{y}} = D(\mathbf{x})\text{.}
\end{equation}
During the training phase, the model is optimized to minimize the classification loss $\mathcal{L}(\hat{\mathbf{y}}, \mathbf{y})$ between the predictions $\hat{\mathbf{y}}$ and the corresponding ground truth labels $\mathbf{y}$. Note that  $\mathbf{y}$ is one-hot encoded vector.
% $\mathbf{Y} = \{\mathbf{y}_1, \mathbf{y}_2, ..., \mathbf{y}_B\}$.

% In order to increase the variety of fake data, we propose to mixed the batch with data from the original input ($\mathbf{X}^o$ and $\mathbf{Y}^o$) and augmented data ($\mathbf{X}^a$ and $\mathbf{Y}^a$). In this work, we propose to add perturbation $\mathbf{P}$ to the original data to create the augmented data, as follows:
% \begin{equation}
%     \mathbf{X}^a = \mathbf{X}^o + \mathbf{P}\text{.}
% \end{equation}
% Meanwhile, $\mathbf{Y}^a$ is set to fake as in this work, the perturbed data is considered to have different pattern than the real data.
% For each of the batch member ($\mathbf{X}$ and $\mathbf{Y}$), we select a member from the augmented data with probability $p$. Otherwise, we select the member from the original data.

To enhance the diversity of fake data, we propose mixing the training batch with both the original and the augmented data. Let us denote by $\mathbf{x}^o$ a given original input data and $\mathbf y^o$ its corresponding label. In this approach, we introduce a perturbation denoted as $\mathbf{p}$
% $\mathbf{P} = \{\mathbf{p}_1, \mathbf{p}_2, ..., \mathbf{p}_B\}$ 
to the original data for creating augmented data. In other words, given an original data $\mathbf{x}^o$, we obtain an augmented sample denoted as $\mathbf{x}^a$ as follows,
% \begin{equation}
% \mathbf{X}^a = \mathbf{X}^o + \mathbf{P}\text{.}
% \end{equation}
\begin{equation}
\mathbf{x}^a = \mathbf{x}^o + \mathbf{p}\text{.}
\end{equation}
Meanwhile, the corresponding label of $\mathbf{x}^a$ denoted as  $\mathbf{y}^a$ is set to \textit{fake} (i.e., $\mathbf{y}^a = \begin{pmatrix}
           0 \\
           1 \\
         \end{pmatrix}$) since the perturbed data is presumed to exhibit a different pattern compared to real data. From each data in the training batch, we randomly select a member from the augmented data or the original data with a probability of $p \in [0,1]$ and  $1-p$, respectively.

% To target the decision boundary of the model, we propose a targeted perturbation method inspired by adversarial attacks \cite{goodfellow2015explaining}:
% \begin{equation}
%     \mathbf{P} = - \epsilon * \text{sign}(\nabla_\mathbf{X} \mathcal{L}(\hat{\mathbf{Y}^o}, \tilde{\mathbf{Y}}))\text{,}
% \end{equation}
% where $\hat{\mathbf{Y}^o} = D(\mathbf{X}^o)$, $\epsilon$ is perturbation strength randomly selected between hyperparameter minimum value $\epsilon_{min}$ and hyperparameter maximum value $\epsilon_{max}$, and $\tilde{\mathbf{Y}}$ is target set to half real and half fake. Negative sign is to go against the gradient direction (i.e., gradient descent). By targeting half-real-half-fake, we generate data near the real boundary which we believe can improve the model performance better compared to if targeting fake. We illustrate the intuition in Figure \ref{fig:method_intuition}, where our method targeting half-real-half-fake is depicted in Figure \ref{fig:method_intuition}(a) and targeting fake in Figure \ref{fig:method_intuition}(b). 
% It can be noted that, unlike adversarial attacks that only look for the weakness of the model, we utilize the perturbed input as augmented synthetic data to increase the performance of the model. Moreover, we select one of the simplest method in adversarial attack to proof the concept, however, the idea can be transferred to any adversarial attack method. The overall algorithm to generate a training batch is summarized in Algorithm \ref{alg:generatebatch}.

To target the decision boundary of the model, we propose an adaptive perturbation method inspired by adversarial attacks~\cite{goodfellow2015explaining}. The perturbation of a data $\mathbf{p}$ is computed using the gradient of the classification loss with respect to the input data $\nabla_{\mathbf{x}} \mathcal{L}(\hat{\mathbf{y}}^o, \tilde{\mathbf{y}})$ as follows,
\begin{equation}
\mathbf{p} = - \epsilon * \text{sign}(\nabla_{\mathbf{x}} \mathcal{L}(\hat{\mathbf{y}}^o, \tilde{\mathbf{y}}))\text{,}
\label{eq:perturbation}
\end{equation}
where $\hat{\mathbf{y}}^o = D(\mathbf{y}^o)$ represents the prediction probability of the original data, $\epsilon$ denotes the perturbation magnitude which is randomly selected between the minimum and the maximum hyperparameter values $\epsilon_{\text{min}}$ and $\epsilon_{\text{max}}$ that are fixed empirically. The variable $\tilde{\mathbf{y}}$ is set to be ambiguous with equal probability predictions for real and fake classes (i.e., $\tilde{\mathbf{y}}=\begin{pmatrix}
           0.5 \\
           0.5 \\
         \end{pmatrix}$). We assume that following such a strategy will be a way for targeting the decision boundary. The negative sign is utilized to move against the direction of the gradient (i.e., gradient descent).
As depicted in Figure~\ref{fig:method_intuition}(a), generating augmented data in the neighborhood of the decision boundary can  enhance the model performance in detecting unseen fake samples as compared to targeting confident fake prediction (i.e., $\tilde{\mathbf{y}}=\begin{pmatrix}
           0 \\
           1 \\
         \end{pmatrix}$) illustrated in Figure~\ref{fig:method_intuition}(b).
% , we try to generate data near the boundary of real data, which can enhance the model's performance more effectively towards unseen fake compared to targeting fake (i.e., $\tilde{\mathbf{y}}=[0, 1]$) illustrated in Figure~\ref{fig:method_intuition}(b). 
% We illustrate this intuition in Figure~\ref{fig:method_intuition}, where our method targeting the boundary between real and fake is depicted in Figure~\ref{fig:method_intuition}(a), while targeting fake  (i.e., $\tilde{\mathbf{y}}=[0, 1]$) is shown in Figure~\ref{fig:method_intuition}(b). In Figure~\ref{fig:method_intuition}(a), training with the proposed augmented data can lead to decision boundary that is more generalized to unseen fake compared to the augmented data in Figure~\ref{fig:method_intuition}(b). 
Additionally, it is worth noting that, unlike adversarial training that utilizes the perturbed input to increase the robustness of the model against attacks, we utilize the perturbed input as augmented fake data to increase the variety of fake data itself. Moreover, while we employ one of the simplest methods in adversarial attacks to illustrate this concept, the proposed idea can be adapted to any adversarial attack method. 

\section{Experiments}
\label{sec:experiments}
\subsection{Dataset and metrics}
In our experiments, the ASVspoof 2019 logical access (LA)~dataset \cite{todisco2019asvspoof,wang2020asvspoof} is considered. It comprises three subsets, namely, training, development, and evaluation subsets. The training and development sets contain different attacks denoted as A01-A06, while the evaluation set includes unseen attacks referred to as A07-A19.
% comprising three subsets: train, development, and evaluation. The train and development sets contain attacks generated from six spoofing attack algorithms (A01-A06), while the evaluation set includes attacks generated from thirteen algorithms (A07-A19). Therefore, the attacks in the evaluation set are unseen in the train and development sets. 
Additional information can be found in~\cite{wang2020asvspoof}. We employ the development and evaluation sets as validation and testing sets, respectively. We present the obtained results using the minimum normalized tandem detection cost function (min t-DCF)~\cite{kinnunen2018t} and equal error rate (EER). A lower min t-DCF and EER value indicate a higher classification performance.

\subsection{Experimental setup}
% We reuse the open-sourced codes for RawNet2\footnote{\url{https://github.com/asvspoof-challenge/2021/blob/main/LA/Baseline-RawNet2/model.py}} and AASIST\footnote{\url{https://github.com/clovaai/aasist/blob/main/models/AASIST.py}} as our baseline models. However, we reduce AASIST original batch size from 24 to 16 to fit our hardware, while we set batch size 32 for the RawNet2 similarly to the source code. The training is done using the training set with Adam optimizer with learning rate $0.0001$ and weight decay of $0.0001$.
% Validation set is used to select the best validation accuracy across 100 training epochs and save the model for the final test.

We utilize open-sourced code for RawNet2\footnote{\url{https://github.com/asvspoof-challenge/2021/blob/main/LA/Baseline-RawNet2/model.py}}\cite{tak2021end} and AASIST\footnote{\url{https://github.com/clovaai/aasist/blob/main/models/AASIST.py}}\cite{jung2022aasist} as our baseline models. However, we adjust the original batch size of AASIST from 24 to 16 to accommodate our hardware limitations, while we maintain a batch size of 32 for RawNet2 similar to the original source code. The training is conducted using the Adam optimizer \cite{kingma2014adam}, with a learning rate of $0.0001$ and a weight decay of $0.0001$. The validation set is used to select the most accurate model over 100 training epochs that is used for the final testing phase.

\subsection{Ablation study}

% In this subsection, we compare the effect of each component introduced in this works. First, we need to understand our augmented data in comparison to the baseline (i.e., model trained without augmented data). As seen in Table \ref{tab:ablation}, the augmentation method targeting between real and fake is effective in improving the performance both AASIST and RawNet2 baselines. More detail observation on AASIST in different audio deepfake types can also be seen in Table \ref{tab:ablation_detail} demonstrating that our approach improve the performance in majority of deepfake types. For AASIST, we set $p=0.5$, $\epsilon_{min}=0.01$, and $\epsilon_{max}=0.5$. For RawNet2, we set $p=0.7$, $\epsilon_{min}=0.01$, and $\epsilon_{max}=0.7$. Sensitivity of these hyperparameters is discussed in the Section \ref{subsec:hyperparam}.

In this subsection, the impact of each proposed component is analyzed. Firstly, we report the obtained results using the two baselines  without and with the proposed augmentation. As depicted in Table \ref{tab:ablation}, the proposed augmentation which targets samples with ambiguous probability predictions (half real and half fake) effectively enhances the performance of both AASIST and RawNet2 baselines. Additional details regarding the performance obtained for AASIST across different types of audio deepfakes can be found in Table \ref{tab:ablation_detail}, demonstrating the improvement in performance across the majority of deepfake types. For AASIST, we set $p=0.5$, $\epsilon_{\text{min}}=0.01$, and $\epsilon_{\text{max}}=0.5$, while for RawNet2, we set $p=0.7$, $\epsilon_{\text{min}}=0.01$, and $\epsilon_{\text{max}}=0.7$. We discuss the sensitivity to these hyperparameters in Section \ref{subsec:hyperparam}.

% Secondly, in order to know the importance of target in augmentation, we first compare with untargeted augmentation utilizing Gaussian noise as $\mathbf{P}$:
% \begin{equation}
% \mathbf{P} = \mathcal{G}(0, \sigma)\text{,}
% \label{eq:untargeted_perturbation}
% \end{equation}
% where $\sigma$ represents the standard deviation sampled uniformly between predefined $\sigma_{\text{min}}$ and $\sigma_{\text{max}}$. Note that, for fair comparison, we also search for the best hyperparameter for the untargeted augmentations which leads to $p=0.7$, $\sigma_{min}=0.01$, and $\sigma_{max}=1$. As seen in Table \ref{tab:ablation}, untargeted augmented data can increase the variety of fake data which improve the performance compared to the baseline, however, not as good as the targeted augmentation to between real and fake. 
Secondly, to assess the importance of targeting ambiguous pseudo-fakes, we first compare it with untargeted augmentation by utilizing a Gaussian noise such that $\mathbf{p}$ is defined as follows,
\begin{equation}
\mathbf{p} = \mathcal{G}(0, \sigma)\text{,}
\label{eq:untargeted_perturbation}
\end{equation}
where $\sigma$ represents the standard deviation sampled uniformly between two values predefined  denoted as $\sigma_{\text{min}}$ and $\sigma_{\text{max}}$. Note that, for a fair comparison, we also optimize the hyperparameters for untargeted augmentations, resulting in $p=0.7$, $\sigma_{\text{min}}=0.01$, and $\sigma_{\text{max}}=1$. As observed in Table~\ref{tab:ablation}, untargeted augmented data can increase the diversity of fake data, leading to an improved performance as compared to the baseline. However, it is not as effective as the proposed targeted augmentation.

Thirdly, we explore the alternative of augmenting data by targeting to a confident fake prediction, as illustrated in Figure~\ref{fig:method_intuition}(b). To achieve this, we 
% Thirdly, we explore the alternative of augmenting data by targeting standard pseudo-fakes, as illustrated in Figure~\ref{fig:method_intuition}(b). To achieve this, we perturb the data in order to obtain standard pseudo-fakes. This means that we 
set $\tilde{\mathbf{y}}$ in Eq.~\eqref{eq:perturbation} to the fake category (i.e., $\tilde{\mathbf{y}}=\begin{pmatrix}
           0 \\
           1 \\
         \end{pmatrix}$). For a fair comparison, we optimize the hyperparameters for this setup and find $p=0.3$, $\epsilon_{\text{min}}=0.01$, and $\epsilon_{\text{max}}=0.7$. As observed in Table~\ref{tab:ablation}, augmented data targeting strandard pseudo-fakes can also increase the diversity of fake data, leading, hence improving the performance as compared to the baseline. However, it is not as effective as the proposed augmentation which targets ambiguous augmentation (with equivalent probability predictions for both real and fake classes).

\begin{figure}[]
  \centering
  \includegraphics[width=\linewidth]{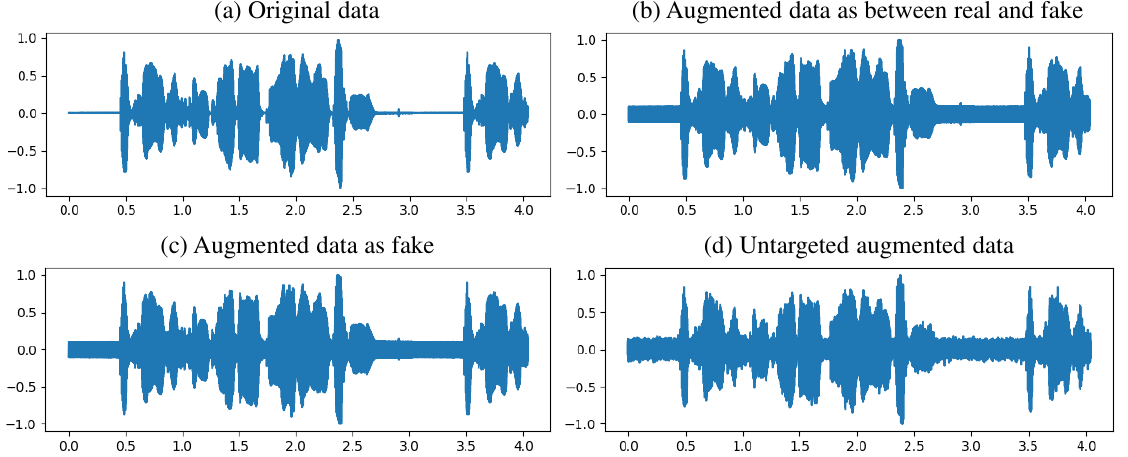}
  \vspace{-5mm}
  \caption{Visualization of an audio sample: (a) without augmentation, (b) with augmentation targeting ambiguous predictions (ours) (c) with augmentation targeting fake predictions, and (d) with untargeted augmentation. Perturbations with approximately similar magnitude are selected for visualization purposes: $\epsilon=0.1$ for (b) and (c), and $\sigma=0.05$ for (d).
  }
  \label{fig:different_augmented_data}
\end{figure}

% In Figure \ref{fig:different_augmented_data}, we visualize the difference between augmented data targeted to between real and fake, augmented data targeted to fake, and untargeted augmented data with similar strength. Interestingly, even though the difference is not very much different visually, the result can still be different.
In Figure \ref{fig:different_augmented_data}, we show qualitative differences among the various augmentation strategies. Interestingly, despite no clear visual distinction can be made, the outcomes can still vary.

\begin{table}[]
\centering
\caption{Ablation study of our method on RawNet2~\cite{tak2021end} and AASIST~\cite{jung2022aasist} baseline models. As observed, the proposed augmentation yields superior results. The best performance in each architecture is marked in bold.}
\resizebox{\linewidth}{!}{
\begin{tabular}{|l|l|c|c|}
\hline
Architecture             & Augmentation (aug.)                                  & Min t-DCF & EER (\%) \\ \hline
\multirow{4}{*}{RawNet2} & No aug.                                & 0.1003  & 3.79     \\
                         & Untargeted aug.                        & 0.0636   & 2.33     \\
                         & Aug. targeting confident fake prediction                  & 0.0688   & 2.41     \\
                         & Aug. targeting  ambiguous prediction (ours) & \textbf{0.0589}   & \textbf{1.98}     \\ \hline
\multirow{2}{*}{AASIST}  & No aug.                                & 0.0596   & 1.90     \\
                         & Aug. targeting  ambiguous prediction (ours)& \textbf{0.0403}   & \textbf{1.39}     \\ \hline
\end{tabular}
}
\label{tab:ablation}
\end{table}

\begin{table*}[]
\centering
\caption{EER (\%)for each type of fake audio in the evaluation set using the AASIST~\cite{jung2022aasist} model without (baseline) and with (ours) the proposed augmentation. Our augmented data enhances the performance of the baseline across the majority of types. The best performance in each fake type is marked in bold.}
\vspace{-2mm}
\begin{tabular}{|l|ccccccccccccc|}
\hline
Method   & A07             & A08             & A09             & A10             & A11             & A12             & A13             & A14             & A15             & A16             & A17             & A18             & A19             \\ \hline
Baseline & 0.4244          & 0.8556          & \textbf{0.0170} & 0.5704          & 0.3667          & 0.5466          & 0.3429          & 0.3090          & 0.4244          & 1.0831          & 2.4039          & 5.4122          & \textbf{0.8964} \\
Ours     & \textbf{0.1800} & \textbf{0.6112} & 0.0238          & \textbf{0.4074} & \textbf{0.1460} & \textbf{0.3429} & \textbf{0.1392} & \textbf{0.1630} & \textbf{0.2275} & \textbf{0.9371} & \textbf{2.2987} & \textbf{4.7026} & 1.2393          \\ \hline
\end{tabular}
\vspace{-3mm}
\label{tab:ablation_detail}
\end{table*}

% \subsubsection{Pretraining vs Joint Training}
% Pretraining maybe already more overfitting compared to joint training. 
% % During joint training, the augmented data is also vary depending on the state of the network.  -- not really... when fine tuning after pretraining, the network state also changes

%  \begin{figure}[]
%   \centering
%   \includegraphics[width=\linewidth]{images/different_strength_augmenteddata.pdf}
%   % \vspace{-3mm}
%   \caption{asdfasdf}
%   \label{fig:different_strength_augmenteddata}
% \end{figure}

\subsection{Comparison with SoA}

The comparison with state-of-the-art (SoA) approaches are presented in Table~\ref{tab:soa}. Our method demonstrates competitiveness as compared to most SoA techniques, although it may not achieve the highest performance. Future improvements could involve selecting a more optimal baseline model, as theoretically our augmentation method can be applied with different baselines. Potential directions for future research include exploring this augmentation method with more optimal baselines, such as SENet~\cite{zhang122021effect} and RawGAT-ST~\cite{tak2021endgraph}. Nevertheless, we have demonstrated that our method enhances the performance of AASIST~\cite{jung2022aasist} and RawNet2~\cite{tak2021end}.

% % Please add the following required packages to your document preamble:
% % \usepackage[normalem]{ulem}
% % \useunder{\uline}{\ul}{}
% \begin{table}[]
% \centering
% \caption{asdlkjsaf}
% \resizebox{\linewidth}{!}{
% \begin{tabular}{|l|c|c||l|c|c|}
% \hline
% Method                                  & min t-DCF & EER (\%) & Method                     & min t-DCF       & EER (\%)      \\ \hline
% PC-DARTS \cite{ge2021partially}                & 0.0914    & 4.96     & Res-TSSDNet \cite{hua2021towards} & 0.0481          & 1.64          \\
% SE-Res2Net50 \cite{li2021replay}               & 0.0743    & 2.5      & SENet \cite{zhang122021effect}    & \textit{0.0368} & \textit{1.14} \\
% Resnet18-OC-softmax \cite{zhang2021one}        & 0.059     & 2.19     & RawGAT-ST \cite{tak2021end}       & \textbf{0.0335} & \textbf{1.06} \\
% LCNN-Dual attention \cite{ma2021improved}      & 0.0777    & 2.76     & RawNet2* \cite{tak2021end}        & 0.1003          & 3.79          \\
% ResNet18-LMCL-FM \cite{chen2020generalization} & 0.052     & 1.81     & AASIST* \cite{jung2022aasist}     & 0.0596          & 1.90          \\ \cline{4-6} 
% MCG-Res2Net50 \cite{li2021channel}             & 0.052     & 1.78     & Ours (RawNet2)             & 0.0589          & 1.98          \\
% Raw PC-DARTS \cite{ge2021raw}                  & 0.057     & 1.77     & Ours (AASIST)              & {\underline{0.0452}}    & {\underline{1.54}}    \\ \hline
% \end{tabular}
% }
% \label{tab:soa}
% \end{table}

% \usepackage[normalem]{ulem}
\begin{table}[]
\caption{Comparison of various state-of-the-art (SoA) audio deepfake detection methods on the ASVspoof 2019 dataset. The best, second-best, and third-best performances are indicated by bold, italic, and underline, respectively. The $\dagger$ mark represents our experimental results. Our method demonstrates competitive performance as compared to the SoA approaches.}
\centering
\begin{tabular}{|l|c|c|}
\hline
Method                                  & min t-DCF       & EER (\%)      \\ \hline
PC-DARTS \cite{ge2021partially}                & 0.0914          & 4.96          \\
GMM \cite{tak2020spoofing}                    & 0.0904          & 3.50           \\
ResNet18-GAT-T \cite{tak2021graph}             & 0.0894          & 4.71          \\
SE-Res2Net50 \cite{li2021replay}               & 0.0743          & 2.50           \\
Resnet18-OC-softmax \cite{zhang2021one}        & 0.0590           & 2.19          \\
LCNN-Dual attention \cite{ma2021improved}      & 0.0777          & 2.76          \\
ResNet18-LMCL-FM \cite{chen2020generalization} & 0.0520           & 1.81          \\
MCG-Res2Net50 \cite{li2021channel}             & 0.0520          & 1.78          \\
Raw PC-DARTS \cite{ge2021raw}                  & 0.0517           & 1.77          \\
Res-TSSDNet \cite{hua2021towards}             & 0.0481          & 1.64          \\
SENet \cite{zhang122021effect}                 & \textit{0.0368} & \textit{1.14} \\
RawGAT-ST \cite{tak2021endgraph}                   & \textbf{0.0335} & \textbf{1.06} \\
RawNet2$\dagger$ \cite{tak2021end}                     & 0.1003          & 3.79          \\
AASIST$\dagger$ \cite{jung2022aasist}                  & 0.0596          & 1.90          \\ \hline
Ours (RawNet2)                          & 0.0589          & 1.98          \\
Ours (AASIST)                           & {\underline{0.0403}}    & {\underline{1.39}}    \\ \hline
\end{tabular}
\label{tab:soa}
\end{table}

\subsection{Hyperparameter robustness}
\label{subsec:hyperparam}

Our method is based on three hyperparameters: the probability of using augmented data $p$, the maximum perturbation strength $\epsilon_{\text{max}}$, and the minimum perturbation strength $\epsilon_{\text{min}}$. In this subsection, we analyze the impact of these hyperparameters on the performance of the proposed approach. 
% The experiments are conducted using the RawNet2 architecture.

\begin{figure*}[]
  \centering
  \includegraphics[width=\linewidth]{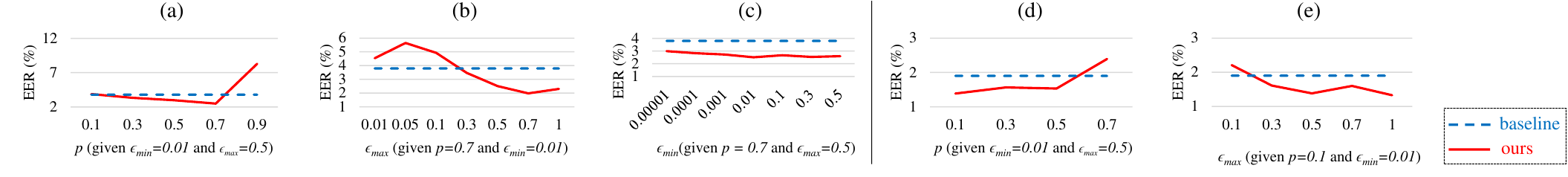}
  \vspace{-7mm}
  \caption{The EER (\%) under different settings. The blue dotted line represents the baseline, i.e., the model trained without augmentation, while the red solid line depicts the model trained with our proposed augmentation technique. Various ranges of hyperparameter values are explored: (a) the probability of augmented data $p$ on RawNet2 (with fixed $\epsilon_{\text{min}}=0.01$ and $\epsilon_{\text{max}}=0.5$), (b) the maximum perturbation strength $\epsilon_{\text{max}}$ on RawNet2 (with fixed $p=0.7$ and $\epsilon_{\text{min}}=0.01$), (c) the minimum perturbation strength $\epsilon_{\text{min}}$ on RawNet2 (with fixed $p=0.7$ and $\epsilon_{\text{max}}=0.5$), (d) the probability of augmented data $p$ on AASIST (with fixed $\epsilon_{\text{min}}=0.01$ and $\epsilon_{\text{max}}=0.5$), and (e) the maximum perturbation strength $\epsilon_{\text{max}}$ on AASIST (with fixed $p=0.1$ and $\epsilon_{\text{min}}=0.01$). Lower EER values indicate better performance.}
  \vspace{-3mm}\label{fig:hyperparameter_eval}
\end{figure*}

% The exploration of $p$ values can be seen in Figure \ref{fig:hyperparameter_eval}(a). As seen, too high $p$, e.g., $p=0.9$ can be problematic to the model. As the real original data is converted to fake data when augmented, the data imbalance between real and fake can also magnified with our augmentation technique which can lead to the performance decrease with too high $p$.
The exploration of $p$ values is depicted in Figure~\ref{fig:hyperparameter_eval}(a) and (d). It is evident that  high values of $p$, such as $p=0.9$ in RawNet2 or $p=0.7$ in AASIST, can pose challenges to the model. With augmented data potentially converting real original data to fake data, there is a risk of aggravating the data imbalance between real and fake, which may result in decreased performance when excessively increasing $p$.

\begin{figure*}[]
  \centering
  \includegraphics[width=\linewidth]{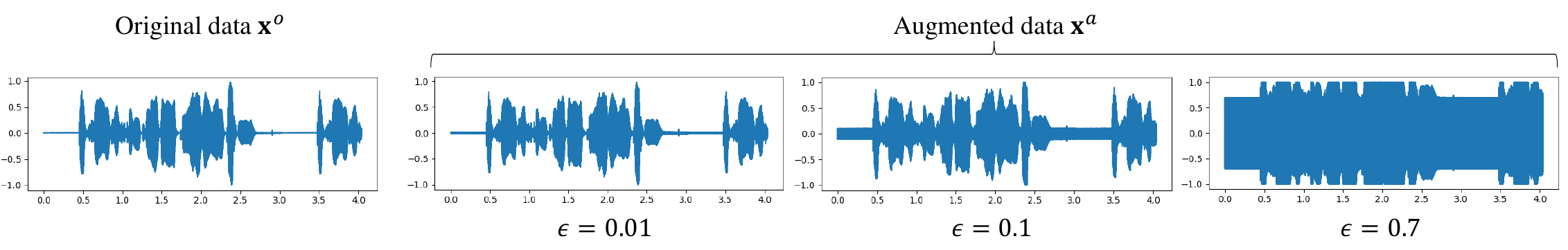}
  \vspace{-7mm}
  \caption{Samples of augmented data with varying perturbation magnitude. Augmentation with a higher magnitude results in augmented data that are more distinct from the original sample.}
  \vspace{-5mm}  \label{fig:different_strength_augmenteddata}
\end{figure*}

% Different perturbation strength can affect how far the augmented data pattern compared to the original data as qualitatively seen in Figure \ref{fig:different_strength_augmenteddata}. As seen in Figure \ref{fig:hyperparameter_eval}(b), setting $\epsilon_{max}$ too small, e.g., $\epsilon_{max} \leq 0.1$ can be harmful to the model as there can be too many augmented data that is too similar to the original data. Especially, when the original data is real, setting the label of similar data to fake can be harmful to the model. 
Different perturbation magnitudes can influence how much the augmented data pattern deviates from the original data, as observed qualitatively in Figure~\ref{fig:different_strength_augmenteddata}. As depicted in Figure~\ref{fig:hyperparameter_eval}(b), setting $\epsilon_{\text{max}}$ too small, for example, $\epsilon_{\text{max}} \leq 0.1$, can be detrimental to the model. This is because there may be an excessive number of augmented data points that closely resemble the original data. This scenario is particularly problematic when the original data is real, as labeling similar data as fake can harm the performance of the model. Similar suited $\epsilon_{\text{max}}$ values can also be used for AASIST as depicted in Figure~\ref{fig:hyperparameter_eval}(e).

% Depending on the robustness of each model, setting $\epsilon_{\text{max}}$ too high can also be detrimental, as illustrated in Figure~\ref{fig:hyperparameter_eval}(e), since it can introduce too many augmented data points that deviate significantly from the original domain.

On the other hand, having a large $\epsilon_{\text{min}}$ is still acceptable for the model, as observed in Figure~\ref{fig:hyperparameter_eval}(c). The model exhibits robustness across a wide range of $\epsilon_{\text{min}}$ values. 
% When the perturbation magnitude is high, the pattern of the augmented data is significantly distinct from the real data, making it safer to be classified as fake data. 
It is important to note that having a different pattern compared to the original data does not necessarily mean that the data lies/does not lie on the decision boundary. An augmented data point can be dissimilar to the original data while still residing in the neighborhood of decision boundary (where the model predicts half-real-half-fake), especially if the boundary is far from the original data.

\section{Conclusion}
\label{sec:conclusion}
In this study, we introduce a novel data synthesis strategy for enhancing the performance of audio deepfake detection methods. Our approach generates fake data by focusing on ambiguous predictions, by taking inspiration from adversarial attack techniques. The proposed augmentation method has the advantage of being architecture-agnostic, hence can be coupled with any audio deepfake detection technique. Through extensive extensive, we demonstrate the relevance of targeting ambiguous augmentations in the context of audio deepfake detection, highlighting the need of further exploring this research area. In future works, we intend to explore more recent adversarial attack techniques.  

\bibliographystyle{IEEEtran}
\bibliography{references}

\begin{thebibliography}{10}
\providecommand{\url}[1]{#1}
\csname url@samestyle\endcsname
\providecommand{\newblock}{\relax}
\providecommand{\bibinfo}[2]{#2}
\providecommand{\BIBentrySTDinterwordspacing}{\spaceskip=0pt\relax}
\providecommand{\BIBentryALTinterwordstretchfactor}{4}
\providecommand{\BIBentryALTinterwordspacing}{\spaceskip=\fontdimen2\font plus
\BIBentryALTinterwordstretchfactor\fontdimen3\font minus \fontdimen4\font\relax}
\providecommand{\BIBforeignlanguage}[2]{{%
\expandafter\ifx\csname l@#1\endcsname\relax
\typeout{** WARNING: IEEEtran.bst: No hyphenation pattern has been}%
\typeout{** loaded for the language `#1'. Using the pattern for}%
\typeout{** the default language instead.}%
\else
\language=\csname l@#1\endcsname
\fi
#2}}
\providecommand{\BIBdecl}{\relax}
\BIBdecl

\bibitem{biden-fake}
J.~Kelety, ``{Fake audio falsely claims to reveal private Biden comments},'' \url{https://apnews.com/article/fact-check-biden-audio-banking-fake-746021122607}, 2023, [Online; accessed 28-February-2024].

\bibitem{scam}
E.~Flitter and S.~Cowley, ``{Voice Deepfakes Are Coming for Your Bank Balance},'' \url{https://www.nytimes.com/2023/08/30/business/voice-deepfakes-bank-scams.html}, 2023, [Online; accessed 28-February-2024].

\bibitem{jung2022aasist}
J.-w. Jung, H.-S. Heo, H.~Tak, H.-j. Shim, J.~S. Chung, B.-J. Lee, H.-J. Yu, and N.~Evans, ``Aasist: Audio anti-spoofing using integrated spectro-temporal graph attention networks,'' in \emph{ICASSP 2022-2022 IEEE International Conference on Acoustics, Speech and Signal Processing (ICASSP)}.\hskip 1em plus 0.5em minus 0.4em\relax IEEE, 2022, pp. 6367--6371.

\bibitem{tak2021endgraph}
H.~Tak, J.~weon Jung, J.~Patino, M.~Kamble, M.~Todisco, and N.~Evans, ``{End-to-end spectro-temporal graph attention networks for speaker verification anti-spoofing and speech deepfake detection},'' in \emph{Proc. 2021 Edition of the Automatic Speaker Verification and Spoofing Countermeasures Challenge}, 2021, pp. 1--8.

\bibitem{tak2021end}
H.~Tak, J.~Patino, M.~Todisco, A.~Nautsch, N.~Evans, and A.~Larcher, ``End-to-end anti-spoofing with rawnet2,'' in \emph{ICASSP 2021-2021 IEEE International Conference on Acoustics, Speech and Signal Processing (ICASSP)}.\hskip 1em plus 0.5em minus 0.4em\relax IEEE, 2021, pp. 6369--6373.

\bibitem{ge2021partially}
W.~Ge, M.~Panariello, J.~Patino, M.~Todisco, and N.~Evans, ``Partially-connected differentiable architecture search for deepfake and spoofing detection,'' in \emph{Interspeech 2021}.\hskip 1em plus 0.5em minus 0.4em\relax ISCA, 2021, pp. 4319--4323.

\bibitem{ge2021raw}
W.~Ge, J.~Patino, M.~Todisco, and N.~Evans, ``Raw differentiable architecture search for speech deepfake and spoofing detection,'' in \emph{ASVSPOOF 2021, Automatic Speaker Verification and Spoofing Countermeasures Challenge}.\hskip 1em plus 0.5em minus 0.4em\relax ISCA, 2021, pp. 22--28.

\bibitem{zaheer2020old}
M.~Z. Zaheer, J.-h. Lee, M.~Astrid, and S.-I. Lee, ``Old is gold: Redefining the adversarially learned one-class classifier training paradigm,'' in \emph{Proceedings of the IEEE/CVF Conference on Computer Vision and Pattern Recognition}, 2020, pp. 14\,183--14\,193.

\bibitem{astrid2021learning}
M.~Astrid, M.~Z. Zaheer, J.-Y. Lee, and S.-I. Lee, ``Learning not to reconstruct anomalies,'' \emph{British Machine Vision Conference}, 2021.

\bibitem{ngo2019fence}
P.~C. Ngo, A.~A. Winarto, C.~K.~L. Kou, S.~Park, F.~Akram, and H.~K. Lee, ``Fence gan: Towards better anomaly detection,'' in \emph{2019 IEEE 31St International Conference on tools with artificial intelligence (ICTAI)}.\hskip 1em plus 0.5em minus 0.4em\relax IEEE, 2019, pp. 141--148.

\bibitem{dionelis2020tail}
N.~Dionelis, M.~Yaghoobi, and S.~A. Tsaftaris, ``Tail of distribution gan (tailgan): Generativeadversarial-network-based boundary formation,'' in \emph{2020 Sensor Signal Processing for Defence Conference (SSPD)}.\hskip 1em plus 0.5em minus 0.4em\relax IEEE, 2020, pp. 1--5.

\bibitem{dionelis2020boundary}
------, ``Boundary of distribution support generator (bdsg): sample generation on the boundary,'' in \emph{2020 IEEE International Conference on Image Processing (ICIP)}.\hskip 1em plus 0.5em minus 0.4em\relax IEEE, 2020, pp. 803--807.

\bibitem{goodfellow2015explaining}
\BIBentryALTinterwordspacing
I.~Goodfellow, J.~Shlens, and C.~Szegedy, ``Explaining and harnessing adversarial examples,'' in \emph{International Conference on Learning Representations}, 2015. [Online]. Available: \url{http://arxiv.org/abs/1412.6572}
\BIBentrySTDinterwordspacing

\bibitem{hua2021towards}
G.~Hua, A.~B.~J. Teoh, and H.~Zhang, ``Towards end-to-end synthetic speech detection,'' \emph{IEEE Signal Processing Letters}, vol.~28, pp. 1265--1269, 2021.

\bibitem{ma2021improved}
X.~Ma, T.~Liang, S.~Zhang, S.~Huang, and L.~He, ``Improved lightcnn with attention modules for asv spoofing detection,'' in \emph{2021 IEEE International Conference on Multimedia and Expo (ICME)}.\hskip 1em plus 0.5em minus 0.4em\relax IEEE, 2021, pp. 1--6.

\bibitem{li2021replay}
X.~Li, N.~Li, C.~Weng, X.~Liu, D.~Su, D.~Yu, and H.~Meng, ``Replay and synthetic speech detection with res2net architecture,'' in \emph{ICASSP 2021-2021 IEEE international conference on acoustics, speech and signal processing (ICASSP)}.\hskip 1em plus 0.5em minus 0.4em\relax IEEE, 2021, pp. 6354--6358.

\bibitem{li2021channel}
X.~Li, X.~Wu, H.~Lu, X.~Liu, and H.~Meng, ``Channel-wise gated res2net: Towards robust detection of synthetic speech attacks,'' \emph{arXiv preprint arXiv:2107.08803}, 2021.

\bibitem{tak2021graph}
H.~Tak, J.-w. Jung, J.~Patino, M.~Todisco, and N.~Evans, ``Graph attention networks for anti-spoofing,'' \emph{arXiv preprint arXiv:2104.03654}, 2021.

\bibitem{zhang122021effect}
Y.~Zhang12, W.~Wang12, and P.~Zhang12, ``The effect of silence and dual-band fusion in anti-spoofing system,'' in \emph{Proc. Interspeech}, 2021.

\bibitem{li2020face}
L.~Li, J.~Bao, T.~Zhang, H.~Yang, D.~Chen, F.~Wen, and B.~Guo, ``Face x-ray for more general face forgery detection,'' in \emph{Proceedings of the IEEE/CVF conference on computer vision and pattern recognition}, 2020, pp. 5001--5010.

\bibitem{shiohara2022detecting}
K.~Shiohara and T.~Yamasaki, ``Detecting deepfakes with self-blended images,'' in \emph{Proceedings of the IEEE/CVF Conference on Computer Vision and Pattern Recognition}, 2022, pp. 18\,720--18\,729.

\bibitem{mejri2023untag}
N.~Mejri, E.~Ghorbel, and D.~Aouada, ``Untag: Learning generic features for unsupervised type-agnostic deepfake detection,'' in \emph{ICASSP 2023-2023 IEEE International Conference on Acoustics, Speech and Signal Processing (ICASSP)}.\hskip 1em plus 0.5em minus 0.4em\relax IEEE, 2023, pp. 1--5.

\bibitem{nguyen2024laa}
D.~Nguyen, N.~Mejri, I.~P. Singh, P.~Kuleshova, M.~Astrid, A.~Kacem, E.~Ghorbel, and D.~Aouada, ``Laa-net: Localized artifact attention network for high-quality deepfakes detection,'' \emph{arXiv preprint arXiv:2401.13856}, 2024.

\bibitem{chen2022self}
L.~Chen, Y.~Zhang, Y.~Song, L.~Liu, and J.~Wang, ``Self-supervised learning of adversarial example: Towards good generalizations for deepfake detection,'' in \emph{Proceedings of the IEEE/CVF conference on computer vision and pattern recognition}, 2022, pp. 18\,710--18\,719.

\bibitem{todisco2019asvspoof}
M.~Todisco, X.~Wang, V.~Vestman, M.~Sahidullah, H.~Delgado, A.~Nautsch, J.~Yamagishi, N.~Evans, T.~Kinnunen, and K.~A. Lee, ``Asvspoof 2019: Future horizons in spoofed and fake audio detection,'' in \emph{INTERSPEECH 2019-20th Annual Conference of the International Speech Communication Association}, 2019.

\bibitem{wang2020asvspoof}
X.~Wang, J.~Yamagishi, M.~Todisco, H.~Delgado, A.~Nautsch, N.~Evans, M.~Sahidullah, V.~Vestman, T.~Kinnunen, K.~A. Lee \emph{et~al.}, ``Asvspoof 2019: A large-scale public database of synthesized, converted and replayed speech,'' \emph{Computer Speech \& Language}, vol.~64, p. 101114, 2020.

\bibitem{kinnunen2018t}
T.~Kinnunen, K.~A. Lee, H.~Delgado, N.~Evans, M.~Todisco, M.~Sahidullah, J.~Yamagishi, and D.~A. Reynolds, ``t-dcf: a detection cost function for the tandem assessment of spoofing countermeasures and automatic speaker verification,'' in \emph{The Speaker and Language Recognition Workshop (Odyssey 2018)}.\hskip 1em plus 0.5em minus 0.4em\relax ISCA, 2018.

\bibitem{kingma2014adam}
D.~P. Kingma and J.~Ba, ``Adam: A method for stochastic optimization,'' \emph{arXiv preprint arXiv:1412.6980}, 2014.

\bibitem{tak2020spoofing}
H.~Tak, J.~Patino, A.~Nautsch, N.~Evans, and M.~Todisco, ``Spoofing attack detection using the non-linear fusion of sub-band classifiers,'' in \emph{Interspeech 2020}.\hskip 1em plus 0.5em minus 0.4em\relax ISCA, 2020, pp. 1106--1110.

\bibitem{zhang2021one}
Y.~Zhang, F.~Jiang, and Z.~Duan, ``One-class learning towards synthetic voice spoofing detection,'' \emph{IEEE Signal Processing Letters}, vol.~28, pp. 937--941, 2021.

\bibitem{chen2020generalization}
T.~Chen, A.~Kumar, P.~Nagarsheth, G.~Sivaraman, and E.~Khoury, ``Generalization of audio deepfake detection.'' in \emph{Odyssey}, 2020, pp. 132--137.

\end{thebibliography}

\end{document}